\documentclass[12pt]{iopart}

\usepackage{hyperref}
\usepackage{graphicx}

\begin{document}

\newcommand{\ket}[1]{\ensuremath{\left| #1 \right>}}
\newcommand{\bra}[1]{\ensuremath{\left< #1 \right|}}
\newcommand{\atstate}[3]{#1#2$_{#3}$}

\title{Quantum Computing Using Crossed Atomic Beams}
\author{P Blythe and B Varcoe}
\address{Department of Physics, University of Sussex, Falmer, Brighton, BN1 9QH, UK}
\ead{p.blythe@sussex.ac.uk}
\date{\today}
\begin{abstract}
A quantum computer is a hypothetical device in which the laws of quantum mechanics are used to introduce a degree of parallelism into computations and which could therefore significantly improve on the computational speed of a classical computer at certain tasks. Cluster state quantum computing (recently proposed by Raussendorf and Briegel) is a new paradigm in quantum information processing and is a departure from the conventional model of quantum computation. The cluster state quantum computer begins by creating a highly entangled multi-particle state (the cluster state) which it uses as a quantum resource during the computation. Information is processed in the computer via selected measurements on individual qubits that form the cluster state. We describe in detail how a scalable quantum computer can be constructed using microwave cavity QED and, in a departure from the traditional understanding of a computer as a fixed array of computational elements, we show that cluster state quantum computing is well suited to atomic beam experiments. We show that all of the necessary elements have been individually realised, and that the construction of a truly scalable atomic beam quantum computer may be an experimental reality in the near future.   
\end{abstract}

\maketitle

Quantum computing presents us with a method of performing parallel calculations using the additional degrees of freedom available by exploiting the entanglement of arbitrary superpositions of computational states. This can give a quantum computer a significant advantage over a classical computer in performing certain tasks. Since its introduction \cite{Deutsch1985} , the field of quantum information processing has made remarkable progress, with recent demonstrations \cite{Roos2004, Leibfried2004, Stick2006} representing the rich potential of the experimental field. While many systems are in principle capable of performing quantum gate operations, some work remains before the ability of these systems to produce a scalable quantum computer is fully understood \cite{Spiller2005}. Recently a new model was introduced \cite{Raussendorf2003} which performs computations based on measurements of qubits in a highly entangled state called a cluster state. The quantum gate operations occur as a sequence of measurements performed on neighbouring qubits, each in a selected basis \cite{Raussendorf2003, Jozsa2005, Nielsen2003, Tame2005, Kiesel2005, Walther2005a, Zhao2004} using feed-forward of measurement outcomes to influence the path of the computation and compensate for the inherent randomness of the measurements. The initial cluster state of qubits is treated as an entanglement resource to be exploited by a computation. The physical qubits of the entangled array act as the carriers of quantum information and once measured play no further role in the computation. The hardest part of quantum computing, the generation of entanglement, is completed before the computation starts, and therefore many of the normal sources of decoherence do not affect the subsequent computation, making the computer more robust. Moreover, any quantum logic operation can be deterministically carried out by a suitable set of measurements on a sufficiently large cluster state \cite{Jozsa2005}. It has also been shown that any polynomial-sized quantum gate array can be implemented in a cluster state quantum computer using at most a polynomial number of measurement layers \cite{Jozsa2005} and is therefore capable of performing certain tasks exponentially faster than a classical computer.

The benefit of using cluster states is that quantum computing may be performed with devices that are capable of generating an entanglement resource in their primary mode of operation. Nevertheless, decoherence of this array remains an issue and we require a device with a long coherence time in comparison with other time scales in the system. An electromagnetic quantum bus \cite{Spiller2006}, such as the quantised field of a high-Q cavity, can be used as the mediator of atom-atom interactions in an atomic beam, thereby turning an atomic beam into a continuous entanglement resource. This attractive feature is the primary reason for the choice of an atomic-beam computing method in this proposal. Other examples of systems potentially capable of cluster-state generation include two-photon down conversion \cite{Kiesel2005, Walther2005a, Zhao2004}, which has recently been used to perform quantum computing gate operations, cavity QED \cite{Zou2005, Hagley1997, Englert2000, Englert1998, Cho2005, Ye2005}, neutral atoms trapped in optical lattices \cite{Kok2005, Mandel2003, Palmer2005, Pachos2003}, linear optics \cite{Hutchinson2004, Nielsen2004, Lim2005, Browne2005, Barrett2005}, and solid state systems \cite{Weinstein2005}. Scalability is an important feature of any new computational scheme and in cluster state quantum computing the ultimate size of the processor is limited by the size of the initial entangled array as this forms the computational working space. Cluster state gate operations use many lattice qubits to perform operations that would require many fewer qubits in other approaches to quantum information processing and the ability to perform qubit-qubit interactions on a vast scale is a central issue.

Microwave cavity QED with Rydberg atoms is a well established field with recent experiments including the creation of an on demand single atom source \cite{Varcoe2004, Brattke2001}, the direct observation of photon Fock states \cite{Maitre1997, Varcoe2000}, the first observation of Fock state coherent population trapping \cite{Weidinger1999}, the observation of non-destructive measurements of a single photon \cite{Nogues1999}, the implementation of controlled phase gates \cite{Rauschenbeutel1999}, controlled atom-atom collisions \cite{Osnaghi2001}, controlled three particle entanglement \cite{Rauschenbeutel2000}, and the two photon micromaser \cite{Brune1987}. Microwave cavity QED delivers long coherence times using superconducting cavities with high $Q$-Factors ($Q \sim 5 \times 10^{10}$) and long decay times ($> 0.3$~s) due to the long wavelengths involved \cite{Varcoe2004, Brattke2001}. Rydberg atoms couple strongly to the cavity field and are in the strong coupling regime of cavity QED. In this regime the interaction is fully coherent, and the evolution of the joint state of the atom and cavity mode is therefore a basic entangling operation. This is used in a number of theoretical papers as a prototype quantum computational device.

Two relatively recent developments aid the realisability of a scalable quantum computer that uses microwave cavity QED with atomic Rydberg transitions. The first is the construction of miniaturised superconducting re-entrant microwave cavities on a millimetre scale \cite{Heerlein1998}. As the processor size scales with the inverse square of the cavity dimension a significant size reduction permits a much more complex system than was previously possible. The second observation is that ionization by tunnelling is a highly efficient method of state selective detection, with efficiencies over 80\% being reported \cite{Tada2002}. 

\begin{figure}[tbp]
	\begin{center}
	\includegraphics{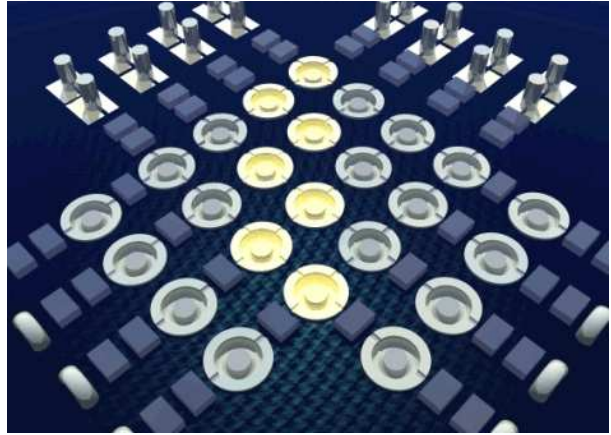}
	\end{center}
	\caption{\label{fig:device}The figure shows a microfabricated array of re-entrant microwave cavities (circular elements), combined with single-qubit rotation zones (rectangular) and state-sensitive detectors (pillars), arranged to form a complete quantum processor. The roles of the microwave cavities are explained in detail in Figure \ref{fig:operation}. The physical dimensions of this chip would be approximately 30mm x 30 mm. The highlighted-yellow collision cavities are sufficient to produce a 2D cluster state, demonstrating that the size of the processor scales linearly with the number of parallel qubit beams. The additional cavities can be turned on or off to allow other topologies of entanglement, forming graph states, increasing the mobility of logical qubit states across the cluster or allowing for the creation of higher dimensional entanglement.}
\end{figure}

Figure \ref{fig:device} presents an impression of a microfabricated array of miniaturised microwave cavities (the circular elements) Ramsey zones (the rectangular elements) and detectors (the pillars) that can be constructed with current state of the art manufacturing facilites \cite{Taylor2005-PrivComm}. The elements will be fabricated on a chip with a specialist high speed surface micro-milling technique on thin films of niobium or aluminium that have been deposited on a substrate by cold spraying \cite{Taylor2005-PrivComm}. The basic physical dimensions of these devices are well defined and this should give a large degree of repeatability in coupling strengths and resonant frequencies across multiple copies of the same cavity design. This makes scaling the production from a single cavity to a large array an entirely realistic proposition.

Building on previously demonstrated experimental results in microwave cavity QED and the recent development of cluster-state quantum computing, in this paper we present an architecture for a scalable, reconfigurable, cavity-QED based cluster state quantum computer.

\section{Cluster state quantum computing}
A cluster state quantum computer uses a sequence of measurements on a lattice of entangled qubits called a cluster state to perform quantum gate operations. The operation of a cluster state quantum computer can be broken down into distinct phases. Firstly, the cluster state is formed when a selection of qubits prepared in the state $\ket{+} = ( \ket{0} + \ket{1} ) / \sqrt(2)$  are entangled in a two dimensional lattice by applying the controlled phase gate $CPhase \ket{+}_1 \ket{+}_2 = \ket{1}_1 \ket{+}_2 + \ket{0}_1 \ket{-}_2$  between pairs of neighbouring lattice qubits (where $ \left\{ \ket{1}_1 \ket{+}_2, \ket{0}_1 \ket{-}_2, \ket{0}_1 \ket{+}_2, \ket{1}_1 \ket{-}_2 \right\}$  is the Schmidt basis and $\ket{-} = ( \ket{0} - \ket{1} ) / \sqrt(2)$ ). The calculation then progresses by making measurements of the lattice qubits on the basis $B(\alpha) = \{ \ket{+ \alpha}, \ket{- \alpha} \}$  (where $\ket{\pm \alpha} = (\ket{0} \pm e^{i \alpha} \ket{1})/ \sqrt(2)$) followed by a ``feed-forward'' rotation of the basis on which subsequent measurements are made conditioned on the measurement outcomes \cite{Raussendorf2003, Jozsa2005, Nielsen2003, Tame2005, Kiesel2005, Walther2005a, Zhao2004}. Two special cases of are measurements in the basis states $\ket{\alpha=0}$ (an X basis measurement) and $\ket{\alpha = \pm \pi/2}$ (a Y basis measurement). Any qubits playing no role in the calculation can be disentangled using a Z-basis measurement, which disentangles qubits from the cluster without affecting the remaining atoms.

\section{The architecture}
Cavity QED is widely believed to be an excellent system for quantum gate operations and recently a number of papers have been published with methods for creating cluster states in equivalent systems \cite{Zou2005, Cho2005, Ye2005}. Any of these methods would be suitable for the creation of linear cluster states in this proposal. The attractive feature of the method we outline here is its passive nature, the low experimental overhead and the high degree of integration that we can achieve.

\begin{figure}[tbp]
	\begin{center}
	\includegraphics{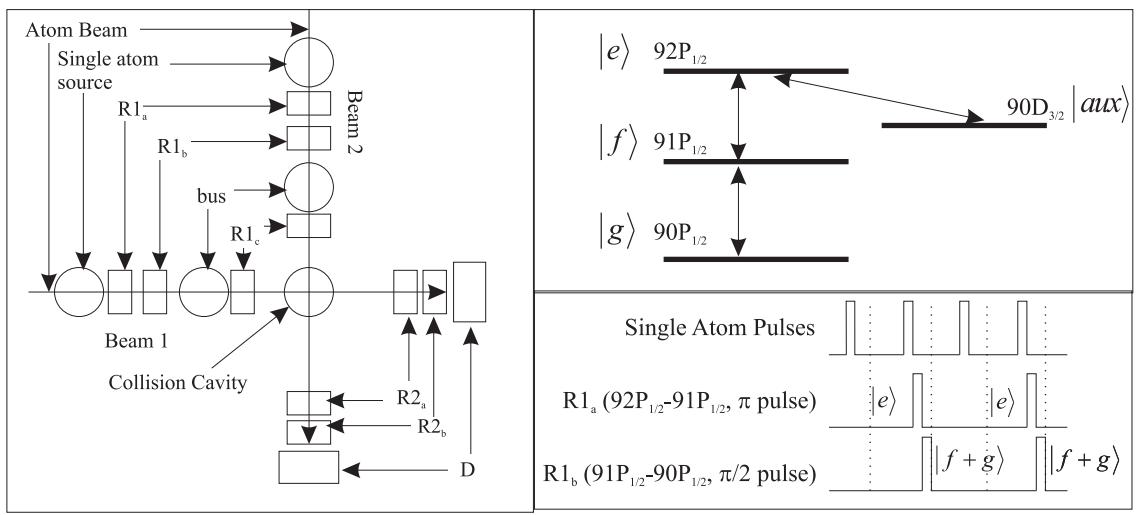}
	\end{center}
	\caption{\label{fig:operation}The basic experimental apparatus, atomic transitions and timing pulses for the creation of linear and two dimensional cluster states. The basic two dimensional entanglement elements consist of two sources of linear cluster states (beams 1 and 2) incident on a collision cavity. The atoms are timed to collide in this cavity thereby turning the linear entangled states into a two dimensional cluster state. The single qubit rotation zones R1a and R1b produce the Hadamard transformation on the input qubits as required. The single qubit rotation zones R2a and R2b can be controlled to produce any arbitrary rotation triggered by the state selective detectors, D on either transition. The lower right hand side shows the sequence of pulses for producing the atomic sequence Hadamard transformation H1 on alternate atoms. The zone R1a can be used to introduce single qubit rotations to atoms in the state.}
\end{figure}

The experimental arrangement in figure \ref{fig:operation} represents a subset of the larger device of figure \ref{fig:device}, encompassing just two atomic beams. The three-level atomic structure and the single qubit rotation sequence for creating the atomic beams is shown. Atoms from the two beams collide in the central cavity, and this collision can be used to produce entanglement between pairs of atoms from perpendicular beams.

To perform the cross beam entanglement, the cavity is used in a dispersive regime by detuning the cavity from resonance by an amount $\delta$  from the  $\ket{e} \leftrightarrow \ket{f}$ transition, but remaining far off-resonance from the $\ket{g} \leftrightarrow \ket{f}$  and $\ket{e} \leftrightarrow \ket{g}$  transitions. The detuning is chosen so that  $\delta \gg \Omega$, where $\Omega$ is the coupling strength between the  $\ket{e} \leftrightarrow \ket{f}$ transition and the cavity mode. Under these circumstances, the following interactions occur:
\begin{eqnarray}
\ket{e}_1 \ket{f}_2 & \rightarrow & e^{-i \lambda t} \left[ \cos{\lambda t} \ket{e}_1 \ket{f}_2 - i \sin{\lambda t} \ket{f}_1 \ket{e}_2 \right] \\
\ket{e}_1 \ket{g}_2 & \rightarrow & e^{-i \lambda t} \ket{e}_1 \ket{g}_2 \\
\ket{f}_1 \ket{f}_2 & \rightarrow & \ket{f}_1 \ket{f}_2 \\
\ket{f}_1 \ket{g}_2 & \rightarrow & \ket{f}_1 \ket{g}_2 \\
\end{eqnarray}

Where $\lambda = \Omega^2 / \delta$. We choose the basis states of the operation to be $\ket{e}_1,\ket{f}_1$ for the first atom and  $\ket{f}_2, \ket{g}_2$ for the second. When $\lambda t = \pi$ the interaction produces a logical $CPhase$ gate \cite{Jozsa2005, Zheng2000}, and this is the basic building-block operation for cluster state creation. The interaction has been experimentally demonstrated with good fidelity by Osnaghi et al \cite{Osnaghi2001}, and the theoretical fidelity of the gate remains high in the presence of small timing errors, achieving 99\% fidelity for a 1\% difference in arrival time relative to interaction time \cite{Zheng2000}.
The interactions occur between one atom prepared in some superposition of the states \ket{e} and \ket{f} and a second atom in the \ket{f} and \ket{g}  states. The atomic sources produce atoms in each basis alternately. Therefore if source B is 'delayed' by one atom relative to source A, then when any two atoms meet in the collision cavity the interaction will occur between atoms prepared in a superposition of the correct basis states. 
The arrangement of the entanglement cavities in the full-scale device (figure \ref{fig:device}) allows the generation of linear cluster states by collisional entanglement. If collisions occur in the highlighted cavities, a linear cluster state is formed, the odd-numbered atoms travelling perpendicular to the even-numbered atoms. Such collisions can be guaranteed if the whole device runs on a clocked scheme, with the atomic sources farther from the origin being delayed by a small amount relative to those nearer, to allow for the time of flight of atoms between the entanglement cavities. The size of this linear cluster state is limited only by the number of atomic beams in the apparatus, and states of this size can be produced in every clock cycle. 

To gain full benefit from the quantum cluster computing model, a state with two-dimensional entanglement is required. To produce a two dimensional state from the above architecture, a 'memory' of some sort must be introduced such that successive atoms from each source are entangled with each other. Tripartite entanglement has been experimentally produced in a cavity-QED system \cite{Rauschenbeutel2000}, and this method integrates well with the collisional entanglement scheme.

The generation of entanglement between clock cycles (and therefore the continuous generation of a 2D cluster state) is performed using the classical microwave fields and micromaser cavities in each beamline, prior to the collisional entanglement. An initial atom A1, prepared in the state \ket{f} interacts with the cavity, which is resonant on the $\ket{e} \leftrightarrow \ket{f}$ transition, producing the state  $\psi_1 = (1 / \sqrt{2})(\ket{f}_1 \ket{0}_C + \ket{e}_1 \ket{1}_C)$ - an entangled state of the atom and the cavity photon number. The second atom A2 to interact with the cavity is prepared in the state $(1 / \sqrt{2} )(\ket{f}_2 + \ket{g}_2)$  by the classical fields R1a and R1b. The interaction time with the cavity is chosen to perform a $CPhase$ gate between the atomic state and the cavity photon number. For this choice of initial state, the gate is equivalent to a 'non-demolition' measurement of the cavity photon number, where the phase of the superposition is flipped in the presence of a photon, and remains unaffected otherwise. The resulting A1-A2-Cavity state is
$ \psi_2 = (1/2)[
\ket{f}_1 \ket{f}_2 \ket{0}_C +
\ket{f}_1 \ket{g}_2 \ket{0}_C +
\ket{e}_1 \ket{f}_2 \ket{1}_C -
\ket{e}_1 \ket{g}_2 \ket{1}_C]$. The state of the cavity can be copied onto a third atom, prepared in \ket{f}, which interacts with the cavity so as to completely absorb a photon, if one is present in the cavity. This returns the cavity to its initial state, containing no photons and no trace of its previous interactions. The final atomic state, now completely decoupled from the cavity state, is
$ \psi_3 = (1/2)[
\ket{f}_1 \ket{f}_2 \ket{f}_3 +
\ket{f}_1 \ket{g}_2 \ket{f}_3 +
\ket{e}_1 \ket{f}_2 \ket{e}_3 -
\ket{e}_1 \ket{g}_2 \ket{e}_3]$. This is a three-particle GHZ state, using the logical definitions $\ket{f} \equiv 0, \{ \ket{e}, \ket{g} \} \equiv 1$  .
The atoms produced by this entanglement scheme can continue to the collisional entanglement area of the chip without further manipulation, as they are already defined on the correct bases ($\ket{f}, \ket{e}$  and $\ket{f}, \ket{g}$). Simply delaying adjacent beamlines by one clock cycle relative to each other ensures that collisions always occur between atoms in the correct bases. An extra atom is inserted in each beam, not entangled with its neighbouring atoms, between each pulse of three entangled atoms, to maintain this situation.

\begin{figure}[tbp]
	\begin{center}
	\includegraphics{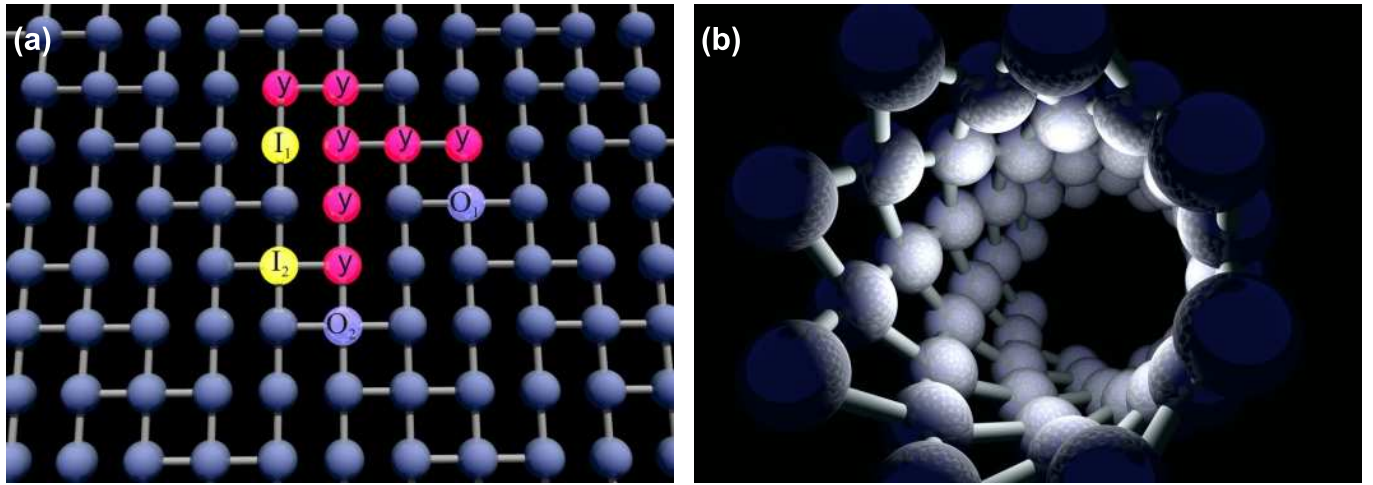}
	\end{center}
	\caption{\label{fig:entanglement}Two types of entanglement structure. The spheres and connections represent qubits and entanglement bonds, respectively. (a) A two dimensional cluster state, produced by the scheme outlined in the paper operating in its default mode.  The coloured qubits show a compact CNOT gate, proceeding from the input qubits (yellow), through a progression of Y-basis measurements (pink), to the output qubits (light blue). In this example the other qubits are all assumed to be measured in Z, and no X-basis measurements, which can be used to propagate logical qubits across the cluster, are used. The size of the state is bounded in the vertical direction by the physical size of the chip, and unbounded in the horizontal direction. (b) Activating additional collision cavities can create higher-dimensional topologies of entanglement like this helical structure.}
\end{figure}

The entangled array of atoms generated by this architecture has the topology shown in figure \ref{fig:entanglement}. It can be seen that in principle computations of essentially arbitrary length can be carried out using such a cluster state, provided that information is routed around the 'missing links'. This consideration does not significantly complicate the design of algorithms for implementation on such a cluster state. For example, a CNOT gate can be efficiently implemented on such a topology, as shown in figure \ref{fig:entanglement}a. The limits to the calculations which can be performed by such a computer will be set by the physical scale of the device (the number of atomic beams), and by the level of coherence which can be practically achieved.

\section{Exploiting the architecture}
An important and powerful feature of the cluster state quantum computer is that it is not necessary for the cluster to be complete before the onset of computational measurements. New qubits can be added to the cluster state as the computation progresses. This is an important feature because it allows atomic beam cluster states that are much larger than the physical site of the apparatus would otherwise allow. 

The highlighted cavities in figure \ref{fig:device} are the only auxiliary cavities required to be in operation to produce a basic two dimensional lattice cluster state; more complicated entanglement geometries can be produced by switching on or off other cavities in this central array. By tuning them to the appropriate off-resonant detuning, they can perform the entangling gate operation required for cluster-state generation. Tuning these cavities into exact resonance can also perform more complex gate operations if required, such as acting as a dynamic photon-based quantum memory, or performing gates which could also involve classical feedback from measurements on earlier qubits \cite{Wiseman1994}. 
The cavity switching can be combined with optical switching of the laser excitation pulses which produce Rydberg atoms, to give complete flexibility over the presence of both individual qubits and any entanglement between them.

This flexibility can be used to generate more complex topologies of entanglement. By turning on an `off-diagonal' cavity in figure \ref{fig:device}, physical qubits at the bottom of the array can be entangled with those at the top. This forms an entangled cluster state with a tubular topology (figure \ref{fig:entanglement}b), around which the logical qubits can move without interruption, allowing computations of (in principle) arbitrary length. By switching other entangling cavities on or off, additional entanglement connections can be generated This allows highly complex topologies to be generated, ranging from tubes, through nested tubes with interconnections, via DNA-like helices with extra cross-links, to regular '3D' cluster states, and many others. This flexibility should allow significant optimisation of algorithms, by simplifying the routing of information across the cluster state, and error correcting trees can also be implemented with little experimental overhead. A future iteration could see wafers stacked vertically linked by atomic beams which generate entanglement between qubits of different wafers. The additional wafers could act as memory or could provide additional entanglement overhead for parallel processing and error correction. 

\section{Implementation}
The microwave cavities on the chip play three roles: as sources of single atoms, as two-photon micromasers generating linear cluster states and as two-atom collisional entanglement cavities for generating two-dimensional cluster states. The cavities are based on a miniature re-entrant design, measuring only a few millimetres across. Such cavities have been demonstrated \cite{Heerlein1998} and $Q$-factors factors as high as $5 \times 10^8$ and coupling strengths (an indication of the gate operation speed) approaching 1~MHz on Rydberg transitions in rubidium can be achieved.

The fabrication of the cavities is the key issue in determining the performance of the system, and this will be an important part of future research. State-of-the-art microfabrication techniques are currently capable of producing the large arrays of cavities, of aluminium or niobium, on a copper surface. However, the microwave characteristics of such cavities have not been thoroughly investigated, and it is expected that significant research on the surface preparation and treatment of the resonators will be necessary to bring them to the performance levels obtained in bulk resonators.

Each atomic beam is created by a deterministic source of single rubidium atoms \cite{Varcoe2004, Brattke2001}, so the position and state of every atom in the system is well defined. A suitable single atom source has been demonstrated with a filling efficiency, limited by the operating conditions, of 83\% and can operate up to a predicted 97\% filling efficiency \cite{Brattke2001, Varcoe2004}. This source can operate on the 
\atstate{92}{P}{1/2} $\leftrightarrow$ \atstate{90}{D}{3/2} one photon maser transition in the conditions for one photon trapping \cite{Brattke2001, Varcoe2004, Maitre1997, Varcoe2000, Weidinger1999}.  In these conditions, a single atomic emission event blocks all other atoms from emitting photons in the cavity, and thus one atom in the \atstate{90}{D}{3/2} state is produced per pulse of excited-state atoms. Linear cross-section atomic beams allow all the parallel atomic paths along one side of the chip to be supplied from a single source. Compact atomic beams with the appropriate properties have been demonstrated \cite{Roach2004}.

The micromaser cavities which generate the entanglement between clock cycles can operate on a one-photon (e.g. \atstate{90}{D}{3/2} $\rightarrow$ \atstate{89}{P}{1/2}) or two-photon (e.g. \atstate{89}{P}{1/2} $\rightarrow$ \atstate{90}{P}{1/2}) transition, with the atoms being prepared in the correct initial state by classical microwave fields. The advantage to working with a one-photon transition is the higher coupling strength, and the availability of proven cavity designs. The advantage offered by a two-photon transition is the potential for lower decoherence induced by stray fields \cite{Peters1991}. Relative shifts of the energy levels induced by external electric or magnetic fields are reduced due to the common quantum numbers (except $n$) of the two states, and the use of $m = 0$ to $m = 0$ transitions, possible for a two-photon transition, can further reduce relative shifts. 

The two photon micromaser has been demonstrated \cite{Brune1987} operating just short of the one-atom-at-a-time regime, and the experiment is made easier by moving to high-$n$ Rydberg transitions. Two factors significantly improve the ability to create a two photon maser: atom-field coupling strength scales as $n^4$, thus an $n = 90$ state is coupled 25 times more strongly than the $n = 40$ states used in previous two-photon maser experiments \cite{Brune1987} and the transition frequency is also significantly reduced making the cavity decay time 6 times longer for the same cavity Q-factor. Moreover although they are more closely separated in energy, Rydberg atoms with $n > 90$ can be state-sensitively detected using tunnelling field ionisation, which has been demonstrated with quantum efficiencies above 80\%, and with an ionisation efficiency above 98\% \cite{Tada2002}.

In addition to the fabrication of the cavities themselves, tuning of the resonant frequencies of the cavities, on an individual basis, is required. For full flexibility, fast switching of the frequencies (to turn cavities 'on' or 'off') is desirable. A general method for such switching is to tune the resonant frequencies of the cavities. This can be achieved for microfabricated cavities by an array of piezoelectric elements mounted to the computational wafer. By changing applied electric potentials on each piezoelectric element, and thus the force applied to each cavity, the physical dimensions of the cavities can be varied, and with them their resonant frequencies. This method is known to work well for larger microwave cavities, and should allow fast (kilohertz speeds) switching of all the cavity elements on the chip. 

\section{Performance and decoherence}
Potential sources of decoherence include the decay lifetime of the microwave cavities, variation of atomic velocities, and the accuracy of the two-atom overlap in the collisional cavities.

The fidelity of the three-atom entangled state from the micromaser will be affected by the uncertainty in the atomic velocities, and therefore interaction times. A velocity resolution of 0.5\% has been achieved in similar experiments \cite{Rauschenbeutel1999}. This limits the fidelity of the three-atom entangled states generated by the system to approximately 99.5\%.

Velocity selection uncertainties have a larger effect on the collisional entanglement process, as they affect not just the transit time of the atoms through the cavity, but also their arrival times, altering the two-atom overlap time. The arrival times of atoms are jittered further due to the mechanism by which the single atom sources operate. A velocity-selected pulse of atoms enters the source cavity, and each atom has a probability $P_e \simeq 0.9$ of emitting a photon into the cavity mode and emerging in the correct atomic state for the subsequent microwave fields. Once this has occurred, the cavity photon blocks the occurrence of further emission events. The time from the start of the atomic pulse to the emission event is then 1.1 atom-transit times along the short axis of the source cavity -- about 0.01~$\mu$s -- with a standard deviation of around   0.01~$\mu$s. In the collisional cavity, the atoms travel along the 'long' axes of the cavity, and so 0.01~$\mu$s  corresponds to 1\% of the atomic overlap time. This is expected to limit the collisional entanglement fidelity to about 99\%, while the differing transit times of the two atoms will contribute to dephasing at below the 1\% level.

Other contributions to dephasing come from stray, fluctuating electric and magnetic fields in the apparatus altering the energies of the atomic states. These fields come from contact potentials due to deposition of atomic vapours on the surfaces of the apparatus, and the grain boundaries in the niobium used to construct the microwave cavities. These effects can be minimised by enclosing the atomic paths with superconducting material, essentially eliminating the stray fields along the beam path, and by preparation and treatment of the niobium cavities to minimise the effects of grain boundaries.

Extrapolating from previously realised microwave cavity designs and known data for lower Rydberg states, the interaction time of an atom with the two-photon micromaser cavity is expected to be on the order of 1~$\mu$s, the time in-between atoms to be around 10~$\mu$s, and the cavity decay lifetime 10~ms. The loss of phase coherence between two successive atoms caused by the finite cavity lifetime would therefore be just 0.05\%. 

There are other small uncertainties associated with the operation of the device, such as the precision of the state rotations, but these can be controlled to a very high level, below a 0.1\% contribution to decoherence, with minimal experimental overhead, and are purely technologically limited.

It can be seen that the cavity decay will not be the limiting factor on the performance of the computer, but other errors such as the overlap of atoms in the collisional entanglement cavity, and small uncertainties in state preparation, mainly due to the accuracy of the selection of atomic velocities, will limit the performance of the gate operations. 

The accuracy of the velocity selection could be improved past the 0.5\% level in several ways. Firstly, the use of longer selection paths, with fast switching tolerances on the microwave fields and careful control of field leakage could improve the resolution. Improved Doppler velocity selection by laser excitation, for example by using selection on all three excitation steps rather than just one, could narrow the distribution a little more. More precise velocity selection schemes are also possible.

The timing jitter of the single atom sources could be improved significantly by the construction of re-entrant microwave cavities with very thin cross-sections, hence low mode volumes and high coupling strengths. Such cavities could also be expected to have higher decay rates, which would allow increases in the rate of atom generation, and hence computational speed.

As a final note, we recognise that scaling the micromaser down to this size and reaching the level of integration required to build a quantum computer presents a formidable technological challenge. However all of the required components of such a system have been previously demonstrated and the techniques for fabricating an integrated device have also been proven in recent years. If all of these strands can be brought together, the result will be a quantum computer with the potential to scale to tens of qubits in the near future, and hundreds or thousands of qubits with refinement of the operating conditions. The system has the potential to cross the threshold for fault-tolerant quantum computing \cite{DiVincenzo1996, Knill2005, Dawson2006}, and therefore scale to arbitrary size.

We have shown that a scalable cavity QED based quantum computer using some of the most recent advances in quantum information science is technically possible. It can utilise error correction schemes with little experimental overhead, is dynamically reconfigurable and uses the well established scientific background of microwave cavity QED.

The authors would like to acknowledge useful discussions with S. Taylor (Liverpool), V. Vedral (Leeds), M. Kim (Dublin), W. Lange (Sussex), W. Munro (Hewlett Packard Laboratories, Bristol) and M. Plenio (Imperial College London). PB is supported by EPSRC grant GR/521892/01.

\bibliographystyle{unsrt}
\bibliography{qc_chip}

\end{document}